\documentclass[12pt,twoside]{article}
\usepackage[eqsecnum]{cmp2e}
\usepackage{citehack}
\usepackage{graphicx}
\usepackage{feynmp,amsbsy,amssymb}

\newcommand{\text}{\mbox}
\title{Strong coupling Hartree-Fock approximation
in the dynamical mean-field theory}
\author{A.M.Shvaika}
\address{Institute for Condensed Matter Physics \\
of the National Academy of Sciences of Ukraine, \\
1~Svientsitskii Str., UA--79011 Lviv, Ukraine}

\begin{document}

\maketitle

\begin{abstract}
In the limit of infinite spatial dimensions a thermodynamically
consistent theory, which is valid for arbitrary value of the Coulombic
interaction ($U<\infty$), is built for the Hubbard model when the total
auxiliary single-site problem exactly splits into four subspaces with
different ``vacuum states''. Some analytical results are given for the
Hartree-Fock approximation when the 4-pole structure for Green's
function is obtained: two poles describe contribution from the Fermi
liquid component, which is ferromagnetic and dominant for small
electron and hole concentrations (``overdoped case'' of high-$T_c$'s),
whereas other two describe contribution from the non-Fermi liquid,
which is antiferromagnetic and dominant close to half filling
(``underdoped case'').
 \pacs 71.10.Fd, 71.15.Mb, 05.30.Fk, 71.27.+a
\end{abstract}

\section{Introduction}

In the last decade the essential achievements in the theory of the
strongly correlated electron systems are connected with the development
of the dynamical mean-field theory (DMFT) proposed by Metzner and
Vollhardt \cite{MetznerVollhardt} for the Hubbard model (see also
Refs.~\cite{IzyumovReview,DMFTreview} and references therein). There
are no restrictions on the $U$ value within this theory and it turns
out to be useful for intermediate coupling ($U\sim t$) for which it
ensures the correct description of the metal-insulator phase transition
and determines the region of the Fermi-liquid behavior of the electron
subsystem. Moreover, some classes of binary-alloy-type models (e.g.,
the Falicov-Kimball model) can be studied almost analytically within
DMFT \cite{BrandtMielsch}. But in the case of the Hubbard model, the
treatment of the effective single impurity Anderson model is very
complicated and mainly computer simulations are used, which calls for
the development of the analytical approaches \cite{Gebhard}.

Such approaches can be built using a systematic perturbation expansion
in terms of the electron hopping \cite{IzyumovCMP} using a diagrammatic
technique for Hubbard operators \cite{Slobodyan,IzyumovBook}. One of
them was proposed for the Hubbard ($U=\infty$ limit) and $t-J$ models
\cite{IzyumovLetfulov}. The lack of such approach is connected with the
concept of a ``hierarchy'' system for the Hubbard operators when the
form of the diagrammatic series and the final results strongly depend
on the system of the pairing priority for Hubbard operators. On the
other hand it is difficult to generalize it to the case of arbitrary
$U$.

In our previous paper \cite{ShvaikaPRB} we developed for the
Hubbard-type models a rigorous perturbation theory scheme in terms of
electron hopping that is based on the Wick theorem for Hubbard
operators \cite{Slobodyan,IzyumovBook}, is valid for arbitrary values
of $U$ ($U<\infty$) and does not depend on the ``hierarchy'' system for
$X$ operators. In the limit of infinite spatial dimensions, this
analytical scheme allows us to build a self-consistent Kadanoff-Baym
type theory \cite{Baym} for the Hubbard model and some analytical
results are given for simple approximations. Here we shall consider
possible magnetic orderings in the Hartree-Fock type approximation.

\section{Perturbation theory in terms of electron hopping}

We consider the lattice electronic system described by the statistical
operator:
\begin{equation}
\label{eq1}
\hat{\rho} = e^{-\beta \hat{H}_{0}}\hat{\sigma}(\beta),\quad
\hat{\sigma}(\beta) = T \!\exp \left\{\!-\!\int\limits_{0}^{\beta} \!\!d\tau
\!\int\limits_{0}^{\beta}\!\!d\tau' \sum\limits_{ij\sigma} t_{ij}^{\sigma}
(\tau-\tau') a_{i\sigma}^{\dag}(\tau) a_{j\sigma}(\tau') \right\},
\end{equation}
where $\hat{H}_{0} = \sum_{i} \hat{H}_{i}$
is a sum of the single-site contributions and for the Hubbard model we must put
\begin{equation}
 H_{i}=Un_{i\uparrow} n_{i\downarrow} - \mu(n_{i\uparrow} +
n_{i\downarrow}) - h(n_{i\uparrow} - n_{i\downarrow}), \qquad
 t_{ij}^{\sigma} (\tau-\tau') = t_{ij} \delta (\tau-\tau').
\label{eq3}
\end{equation}

It is supposed that we know the eigenvalues and eigenstates of the
zero-order Hamiltonian: $H_{i}|i,p\rangle = \lambda_{p} |i,p\rangle$,
and one can introduce Hubbard operators
$\hat{X}_{i}^{pq}=|i,p\rangle\langle i,q|$ in terms of which the
zero-order Hamiltonian is diagonal
\begin{equation}\label{Hdiag}
H_{0}=\sum_{i}\sum_{p} \lambda_{p}\hat{X}_{i}^{pp}.
\end{equation}

For the Hubbard model we have four states
$|i,p\rangle=|i,n_{i\uparrow},n_{i\downarrow}\rangle$:
$|i,0\rangle=|i,0,0\rangle$ (empty site), $|i,2\rangle = |i,1,1\rangle$
(double occupied site), $|i,\uparrow\rangle=|i,1,0\rangle$ and
$|i,\downarrow\rangle=|i,0,1\rangle$ (sites with spin-up and spin-down
electrons) with energies $\lambda_{0}=0$, $\lambda_{2}=U-2\mu$,
$\lambda_{\downarrow} = h-\mu$, and $\lambda_{\uparrow} = -h-\mu$. The
connection between the electron operators and the Hubbard operators is
the following:
\begin{equation}
n_{i\sigma} = X_{i}^{22} + X_{i}^{\sigma\sigma}; \qquad a_{i\sigma} =
X_i^{0\sigma} + \sigma X_i^{\bar\sigma 2}.
\label{eq7}
\end{equation}

The expression for $\langle \sigma(\beta)\rangle_{0}$ is a series of
terms that are products of the hopping integrals and averages of the
electron creation and annihilation operators or Hubbard operators, that
are calculated with the use of the corresponding Wick's theorem
\cite{Slobodyan,IzyumovBook}, and can be written as \cite{ShvaikaPRB}:
\begin{eqnarray}
\left\langle\hat\sigma(\beta)\right\rangle_0&=&\left\langle
\exp\Biggl\{
\right.
-\;
\unitlength=0.1em
\parbox{47\unitlength}{
    \begin{fmffile}{loop1c}
    \begin{fmfgraph}(40,25)\fmfkeep{loopa}
        \fmfpen{thin}\fmfleft{el}\fmfright{er}
        \fmf{photon,right=.7}{er,el}
        \fmf{fermion}{er,el}\fmf{phantom}{er,c1,c2,el}
        \fmf{plain,left,width=thin,tension=.2}{c1,c2,c1}
    \end{fmfgraph}
    \end{fmffile}
}
\!-\frac12\;
\unitlength=0.12em
\parbox{37\unitlength}{
    \begin{fmffile}{loop2c}
    \begin{fmfgraph}(30,25)\fmfkeep{loopb}
        \fmfpen{thin}\fmfleftn{l}{2}\fmfrightn{r}{2}
        \fmf{fermion}{l2,r2}\fmf{phantom}{l2,c1,c2,r2}
        \fmf{plain,left,width=thin,tension=.1}{c1,c2,c1}
        \fmf{photon}{r2,r1}\fmf{fermion}{r1,l1}
        \fmf{phantom}{r1,c3,c4,l1}
        \fmf{plain,left,width=thin,tension=.1}{c3,c4,c3}
        \fmf{photon}{l1,l2}
    \end{fmfgraph}
    \end{fmffile}
}
\!-\frac13\;
\unitlength=0.12em
\parbox{37\unitlength}{
    \begin{fmffile}{loop3c}
    \begin{fmfgraph}(30,25)\fmfkeep{loopc}
        \fmfpen{thin}\fmfleftn{l}{3}\fmfrightn{r}{3}
        \fmftopn{t}{4}\fmfbottomn{b}{4}
        \fmf{fermion}{l2,t2}\fmf{photon}{t2,t3}
        \fmf{fermion}{t3,r2}\fmf{photon}{r2,b3}
        \fmf{fermion}{b3,b2}\fmf{photon}{b2,l2}
        \fmf{phantom}{l2,c1,ca,c2,t2}
        \fmf{phantom}{t3,c3,cb,c4,r2}
        \fmf{phantom}{b3,c5,cc,cd,ce,c6,b2}
        \fmffreeze
        \fmf{plain,left,width=thin}{c1,c2,c1}
        \fmf{plain,left,width=thin}{c3,c4,c3}
        \fmf{plain,left,width=thin}{c5,c6,c5}
    \end{fmfgraph}
    \end{fmffile}
}
\!-\dots
\label{eq17}\\
&&\quad-\;\;
\unitlength=0.16em
\parbox{44\unitlength}{
    \begin{fmffile}{vert2c}
    \begin{fmfgraph*}(40,25)
        \fmfpen{thin}\fmfleftn{l}{3}\fmfrightn{r}{3}
        \fmfpolyn{label=$\bigcirc$}{p}{4}
        \fmf{photon,right=.7}{p1,r2,p2}\fmf{photon,right=.7}{p3,l2,p4}
    \end{fmfgraph*}
    \end{fmffile}
}
\!-\;
\unitlength=0.12em
\parbox{34\unitlength}{
    \begin{fmffile}{vert3c}
    \begin{fmfgraph*}(30,25)
        \fmfpen{thin}\fmftopn{t}{3}\fmfbottomn{b}{2}
        \fmfpolyn{label=$\bigcirc$}{p}{6}
        \fmf{photon,right=.9}{p1,b2,p2}\fmf{photon,right=.9}{p3,t2,p4}
        \fmf{photon,right=.9}{p5,b1,p6}
    \end{fmfgraph*}
    \end{fmffile}
}
\!-\dots-\;
\unitlength=0.16em
\parbox{44\unitlength}{
    \begin{fmffile}{vert2lc}
    \begin{fmfgraph*}(40,15)
        \fmfpen{thin}\fmfleftn{l}{2}\fmfrightn{r}{3}
        \fmfpolyn{label=$\bigcirc$}{p}{4}
        \fmf{photon,right=.7}{p1,r2,p2}\fmf{photon,right=.3}{p3,l2}
        \fmf{fermion}{l1,l2}\fmf{photon,right=.3}{l1,p4}
        \fmf{phantom}{l1,c1,ca,c2,l2}
        \fmf{plain,width=thin,left,tension=.1}{c1,c2,c1}
    \end{fmfgraph*}
    \end{fmffile}
}
\!-\dots
\left.
\Biggr\}\right\rangle_0,
\nonumber
\end{eqnarray}
where arrows denote the zero-order Green's functions
\begin{equation}
g_{pq}(\omega_{n}) = \frac{1}{i\omega_{n}-\lambda_{pq}},
\label{eq16}
\end{equation}
wavy lines denote hopping integrals and $\square$, \ldots\ stand for
complicated ``$n$ vertices''. Each vertex (many-particle single-site
Green's function) is multiplied by a diagonal Hubbard operator denoted
by a circle.

\section{Irreducible many-particle Green's functions}

For the Hubbard model, expressions for the two-vertex
\begin{equation}
\unitlength=0.1em
\parbox{47\unitlength}{\begin{fmffile}{arrow}
    \begin{fmfgraph}(40,10)\fmfkeep{arrow0}
        \fmfpen{thin}\fmfleft{el}\fmfright{er}
        \fmf{fermion}{el,er}\fmf{phantom}{el,c1,c2,er}
        \fmf{plain,left,width=thin,tension=.2}{c1,c2,c1}
    \end{fmfgraph}
    \end{fmffile}}
=\sum_{p} \hat X_{i}^{pp} g_{\sigma(p)}(\omega_{n}),
\label{eq21}
\end{equation}
the four-vertex
\begin{equation}
\label{eq22}
\unitlength=0.1em
\parbox{20\unitlength}{\begin{fmffile}{pure4ver}
\begin{fmfgraph*}(15,15)\fmfkeep{pure4}\fmfstraight\fmfpen{thin}
\fmfleftn{l}{2}\fmfrightn{r}{2}
\fmf{plain}{p1,r1}\fmf{plain}{p2,r2}\fmf{plain}{p3,l2}\fmf{plain}{p4,l1}
\fmfpolyn{label=$\bigcirc$,tension=.2}{p}{4}
\end{fmfgraph*}
\end{fmffile}}=
\sum_{p} \hat{X}_{i}^{pp}
g_{\sigma(p)} (\omega_{n}) g_{\sigma(p)} (\omega_{n+m})
\widetilde{U}_{\sigma\bar{\sigma}(p)} (\omega_{n},\omega_{l}|\omega_{m})
g_{\bar{\sigma}(p)} (\omega_{l}) g_{\bar{\sigma}(p)}
(\omega_{l+m}),
\end{equation}
and for the vertices of higher order possess one significant feature
\cite{ShvaikaPRB}. They decompose into four terms with different
diagonal Hubbard operators $X^{pp}$, which project our single-site
problem on certain ``vacuum'' states (subspaces), and zero-order
Green's functions
\begin{equation}
g_{\sigma(p)}(\omega_{n}) = \left\{ \begin{array}{ll}
g_{\sigma 0} (\omega_{n}) & \text{ for $p=0,\sigma$} \\
g_{2\bar\sigma}(\omega_{n}) & \text{ for $p=\bar\sigma,2$}
\end{array} \right.,
\label{eq23}
\end{equation}
which describe all possible excitations and scattering processes around
these ``vacuum'' states. Here
\begin{equation}
\widetilde{U}_{\sigma\bar{\sigma}(p)} (\omega_{n},\omega_{l}|\omega_{m})
=\left\{ \begin{array} {ll}
\!\!U\pm U^2 g_{20}(\omega_{n+l+m}) & \text{for $p=0,2$} \\
\!\!U\pm U^2 g_{\sigma\bar{\sigma}} (\omega_{n-l}) & \text{for
$p=\sigma,\bar{\sigma}$}
\end{array}
\right.
\label{eq24}
\end{equation}
is a renormalized Coulombic interaction in the subspaces. In diagrammatic
notations expression (\ref{eq22}) can be represented as
\begin{equation}
\unitlength=0.1em\parbox{20\unitlength}
{\begin{fmffile}{pure4vr1}
\begin{fmfgraph}(15,15)\fmfkeep{pure4}\fmfstraight\fmfpen{thin}
\fmfleftn{l}{2}\fmfrightn{r}{2}
\fmf{plain}{p1,r1}\fmf{plain}{p2,r2}\fmf{plain}{p3,l2}\fmf{plain}{p4,l1}
\fmfpolyn{tension=.2}{p}{4}
\end{fmfgraph}
\end{fmffile}}
\;\;=\;\;
\unitlength=0.1em
\parbox{25\unitlength}{\begin{fmffile}{4vert}
\begin{fmfgraph}(20,20)\fmfstraight
    \fmfpen{thin}\fmfleftn{l}{2}\fmfrightn{r}{2}
    \fmf{fermion}{l2,c,l1}\fmf{fermion}{r1,c,r2}\fmfdot{c}
\end{fmfgraph}
\end{fmffile}}
\;\;\pm\;\;
\unitlength=0.1em
\parbox{55\unitlength}{
\begin{fmffile}{4vertd}
\begin{fmfgraph}(50,20)\fmfstraight
    \fmfpen{thin}\fmfleftn{l}{2}\fmfrightn{r}{2}
    \fmf{fermion}{l2,c1}\fmf{fermion}{l1,c1}\fmf{fermion}{c2,r1}
    \fmf{fermion}{c2,r2}\fmf{scalar}{c1,c2}\fmfdotn{c}{2}
\end{fmfgraph}
\end{fmffile}},
\label{eq25}
\end{equation}
where dots denote the Coulombic correlation energy $U$
and the dashed arrows denote bosonic zero-order Green's functions:
doublon $g_{20}(\omega_{m})$ or magnon
$g_{\sigma\bar{\sigma}}(\omega_{m})$. The contributions to the
six-vertex can be presented by the following diagrams:
\begin{equation}
\unitlength=0.1em
\begin{fmffile}{6vert0}
\begin{fmfgraph}(40,20)\fmfstraight
    \fmfpen{thin}\fmfleft{l}\fmfright{r}\fmftopn{t}{5}\fmfbottomn{b}{5}
    \fmf{plain}{l,c1,c2,r}\fmf{plain}{b2,c1,t2}\fmf{plain}{b4,c2,t4}
    \fmfdotn{c}{2}
\end{fmfgraph}
\quad
\begin{fmfgraph}(40,20)\fmfstraight
    \fmfpen{thin}\fmfleft{l}\fmfrightn{r}{4}\fmftopn{t}{5}\fmfbottomn{b}{5}
    \fmf{plain}{l,c1,c2,t3}\fmf{plain}{b2,c1,t2}\fmf{plain}{b4,c3,r2}
    \fmf{dashes}{c2,c3}\fmfdotn{c}{3}
\end{fmfgraph}
\quad
\begin{fmfgraph}(40,20)\fmfstraight
    \fmfpen{thin}\fmfleftn{l}{4}\fmfrightn{r}{4}\fmftopn{t}{6}\fmfbottomn{b}{6}
    \fmf{plain}{l3,c1,t2}\fmf{plain}{b3,c2,c3,b4}\fmf{plain}{t5,c4,r3}
    \fmf{dashes}{c1,c2}\fmf{dashes}{c3,c4}\fmfdotn{c}{4}
\end{fmfgraph}
\quad
\begin{fmfgraph}(40,20)\fmfstraight
    \fmfpen{thin}\fmfleftn{l}{2}\fmfrightn{r}{2}\fmftopn{t}{4}
    \fmf{plain}{l1,c1,l2}\fmf{plain}{r1,c3,r2}
    \fmf{dashes}{c1,c2,c3}\fmfdotn{c}{3}
    \fmffreeze\fmf{plain}{t2,c2,t3}
\end{fmfgraph}
\end{fmffile}
\label{eq26}
\end{equation}
where the first three diagrams contain the internal vertices of the
same type as in Eq.~(\ref{eq25}). So, we can introduce primitive
vertices \raisebox{-.4em}{$ \unitlength=0.07em
\begin{fmffile}{element}
\begin{fmfgraph}(20,20)\fmfstraight
    \fmfpen{thin}\fmfleftn{l}{2}\fmfrightn{r}{2}
    \fmf{plain}{l1,c,l2}\fmf{plain}{r1,c,r2}\fmfdot{c}
\end{fmfgraph}
, \;
\begin{fmfgraph}(20,20)\fmfstraight
    \fmfpen{thin}\fmfleftn{l}{2}\fmfright{r}
    \fmf{plain}{l1,c,l2}\fmf{dashes}{c,r}\fmfdot{c}
\end{fmfgraph}
, \;
\begin{fmfgraph}(20,20)\fmfstraight
    \fmfpen{thin}\fmfleftn{l}{2}\fmfrightn{r}{2}
    \fmf{plain}{l1,c,l2}\fmf{dashes}{r1,c,r2}\fmfdot{c}
\end{fmfgraph}
\end{fmffile}
$} by which one can construct all $n$ vertices in the expansion
(\ref{eq17}).\footnote{For $n$ vertices of higher order a new primitive
vertices can appear but we do not check this due to the rapid increase
of the algebraic calculations with the increase of $n$.}

\section{Dynamical mean-field theory}

Within the framework of the considered perturbation theory in terms of
electron hopping a single-electron Green's function can be presented in
the form
\begin{equation}
G_{\sigma}(\omega_{n},\boldsymbol{k}) = \frac{1}{\Xi_{\sigma}^{-1}
(\omega_{n},\boldsymbol{k}) - t_{\boldsymbol{k}}},
\label{eq29}
\end{equation}
where we introduce an irreducible part
$\Xi_{\sigma}(\omega_{n},\boldsymbol{k})$ of Green's function which, in
general, is not local. In the case of infinite dimensions
$d\rightarrow\infty$ one should scale the hopping integral according to
$t_{ij}\rightarrow {t_{ij}}/{\sqrt{d}}$ in order to obtain finite
density-of-states and it was shown by Metzner in his pioneer work
\cite{Metzner} that in this limit the irreducible part become local
$\Xi_{ij\sigma} (\tau-\tau')= \delta_{ij} \Xi_{\sigma} (\tau-\tau')$ or
$\Xi_{\sigma}(\omega_{n},\boldsymbol{k}) = \Xi_{\sigma} (\omega_{n})$
and such a site-diagonal function can be calculated by mapping the
$d\rightarrow\infty$ lattice problem (\ref{eq1}) onto the atomic model
with the auxiliary Kadanoff-Baym field \cite{BrandtMielsch}
\begin{equation}
t_{ij}^{\sigma}(\tau-\tau') = \delta_{ij} J_{\sigma} (\tau-\tau').
\label{eq31}
\end{equation}
The self-consistent set of equations for
$\Xi_{\sigma}(\omega_{n})$ and $J_{\sigma}(\omega_{n})$
(e.g., see Ref.~\cite{DMFTreview} and references therein) is the following:
\begin{equation}
\frac{1}{N} \sum_{\boldsymbol{k}} \frac{1}{\Xi_{\sigma}^{-1}(\omega_{n})
-t_{\boldsymbol{k}}} = \frac{1}{\Xi_{\sigma}^{-1}(\omega_{n}) -
J_{\sigma}(\omega_{n})}
= G_{\sigma}^{(a)} (\omega_{n}, \{ J_{\sigma}
(\omega_{n})\}),
\label{eq32}
\end{equation}
where $G_{\sigma}^{(a)}(\omega_{n},\{J_{\sigma}(\omega_{n})\})$ is the
Green's function for the atomic limit (\ref{eq31}).

The grand canonical potential for the lattice is connected with the one
for the atomic limit by the expression \cite{BrandtMielsch}
\begin{equation}
\frac{\Omega}{N} = \Omega_{a} - \frac1{\beta} \sum_{n\sigma} \left\{ \ln
G_{\sigma}^{(a)} (\omega_{n}) - \frac{1}{N} \sum_{\boldsymbol{k}} \ln G_{\sigma}
(\omega_{n},\boldsymbol{k})\right\}.
\label{eq33}
\end{equation}

On the other hand, we can write for the grand canonical potential for
the atomic limit $\Omega_{a}$ the same expansion as in Eq.~(\ref{eq17})
but with diagonal $X$ operators at the same site. We can reduce their
product to a single $X$ operator that can be taken outside of the
exponent in (\ref{eq17}) and its average is equal to $\langle
X^{pp}\rangle_{0} = {e^{-\beta\lambda_{p}}}/{\sum_q
e^{-\beta\lambda_{q}}}$. Finally, for the grand canonical potential for
the atomic limit we get \cite{ShvaikaPRB}
\begin{equation}
\Omega_{a} = -\frac{1}{\beta} \ln \sum_{p} e^{-\beta\Omega_{(p)}},
\label{eq34}
\end{equation}
where $\Omega_{(p)}$
are the ``grand canonical potentials'' for the subspaces.

Now we can find the single-electron Green's function for the atomic
limit by
\begin{equation}
G_{\sigma}^{(a)} (\tau-\tau') = \frac{\delta\Omega_{a}}{\delta
J_{\sigma}(\tau-\tau')} = \sum_{p} w_{p} G_{\sigma(p)} (\tau-\tau'),
\label{eq36} \label{eq37}
\end{equation}
where $G_{\sigma(p)} (\tau-\tau')$ are single-electron Green functions
for the subspaces characterized by the ``statistical weights'' $w_{p} =
{e^{-\beta\Omega_{(p)}}}/{\sum_{q}e^{-\beta\Omega_{(q)}}}$ and our
single-site atomic problem exactly splits into four subspaces
$p=0,2,\downarrow,\uparrow$.

The fermionic Green's function in subspaces can be written as
\begin{equation}
G_{\sigma(p)} (\omega_{n}) = \frac{1}{\Xi_{\sigma(p)}^{-1}(\omega_{n}) -
J_{\sigma}(\omega_{n})},
\label{eq39}\quad
\label{GFp}
\Xi_{\sigma(p)}^{-1}(\omega_n)=i\omega_n+\mu_{\sigma}-Un^{(0)}_{\bar\sigma(p)}
-\Sigma_{\sigma(p)}(\omega_n),
\end{equation}
where $n^{(0)}_{\sigma(p)}=-{d\lambda_p}/{d\mu_{\sigma}}=0$ for
$p=0,\bar\sigma$ and $1$ for $p=2,\sigma$ is an occupation of the state
$|p\rangle$ by the electron with spin $\sigma$, and the self-energy
$\Sigma_{\sigma(p)}(\omega_{n})$ depends on the hopping integral
$J_{\sigma'}(\omega_{n'})$ only through quantities
\begin{equation}
\Psi_{\sigma'(p)} (\omega_{n'}) =
G_{\sigma'(p)} (\omega_{n'}) - \Xi _{\sigma'(p)} (\omega_{n'}).
\label{eq41}
\end{equation}

Now, one can reconstruct expressions for the grand canonical potentials
$\Omega_{(p)}$ in subspaces from the known structure of Green's functions:
\begin{equation}
\Omega_{(p)} = \lambda_{p} - \frac{1}{\beta} \sum_{n\sigma}
\ln\left(1 - J_{\sigma}(\omega_{n})\Xi_{\sigma(p)}(\omega_{n})\right)
- \frac{1}{\beta} \sum_{n\sigma} \Sigma_{\sigma(p)}(\omega_{n})
\Psi_{\sigma(p)}(\omega_{n})
+\Phi_{(p)},
\label{eq43}
\end{equation}
where $\Phi_{(p)}$ is some functional, such that its functional
derivative with respect to $\Psi$ produces the self-energy:
${\delta\Phi_{(p)}}/{\delta\Psi_{\sigma(p)}(\omega_{n})}
=\Sigma_{\sigma(p)}(\omega_{n})$. So, if one can find or construct the
self-energy $\Sigma_{\sigma(p)}(\omega_{n})$ he can find Green's
functions and grand canonical potentials for the subspaces and,
according to Eqs.~(\ref{eq34}) and (\ref{eq36}), can solve atomic
problems.

From the grand canonical potential (\ref{eq34}) and (\ref{eq43}) we get
the following mean values
\begin{equation}\label{mv}
n_{\sigma}=\sum_p w_p n_{\sigma(p)}, \quad
n_{\sigma(p)}=n^{(0)}_{\sigma(p)}+
\frac1{\beta}\sum_n\Psi_{\sigma(p)}(\omega_{n})
-\frac{\partial\Phi_{(p)}}{\partial\mu_{\sigma}},
\end{equation}
where in the last term the partial derivative is taken over $\mu_{\sigma}$
not in the chains (\ref{eq41}).

For the Falicov-Kimball model $J_{\downarrow}(\omega_n)=0$ and, as a result,
$\Sigma_{\uparrow(p)}(\omega_{n}) \equiv 0$ and
$\Xi_{\uparrow(p)}(\omega_{n}) = g_{\uparrow(p)} (\omega_{n})$
which immediately gives results of Ref.~\cite{BrandtMielsch}
(see also Ref.~\cite{ShvaikaJPS}).

For the Hubbard model there is no exact expression for the self-energy
but the set of Eqs.~(\ref{eq39}) and (\ref{eq43}) allows one to
construct different self-consistent approximations.

\section{Hartree-Fock approximation}

One of the possible approximations is to construct the equation
for the self-energy in the following form:
\begin{equation}
\Sigma_{\sigma(p)}(\omega_n)=\frac1{\beta}\sum_{n'}
U\Psi_{\bar\sigma(p)}(\omega_{n'}),
\end{equation}
which, together with the expression for mean values
\begin{equation}\label{mvHF}
n_{\sigma(p)} = n^{(0)}_{\sigma(p)} + \frac{1}{\beta} \sum\limits_{n}
\Psi_{\sigma(p)}(\omega_{n}),
\end{equation}
gives for the Green's functions in the subspaces an expression in the
Hartree-Fock approximation:
\begin{equation}\label{GFHF}
G_{\sigma(p)}(\omega_n)=\frac1{i\omega_n+\mu_{\sigma}-Un_{\bar\sigma(p)}
-J_{\sigma}(\omega_n)}.
\end{equation}
Now, the grand canonical potentials in the subspaces are equal
\begin{equation}
\Omega_{(p)}=\lambda_p-\frac1{\beta}\sum_{n\sigma}
\ln\left[1-J_{\sigma}(\omega_n)\Xi_{\sigma(p)}(\omega_n)\right]
-U \left(n_{\sigma(p)}-n_{\sigma(p)}^{(0)}\right)
\left(n_{\bar\sigma(p)}-n_{\bar\sigma(p)}^{(0)}\right)
\end{equation}
and the Green's function for the atomic problem (\ref{eq36}) yield a
four-pole structure, that, in contrast to the alloy-analogy solution
$\Sigma_{\sigma(p)}(\omega_n)=0$, possesses the correct Hartree-Fock
limit for small Coulombic interaction $U\ll t$. On the other hand, in
the same way as an alloy-analogy solution, it describes the
metal-insulator transition with the change of $U$.

In Ref.~\cite{ShvaikaPRB} it was shown that the main contributions into
the total spectral weight function
$\rho_{\sigma}(\omega)=\frac1{\pi}\Im G^{(a)}_{\sigma}(\omega-i0^+)$
come from the subspaces $p=0$ for the low electron concentrations
($n<\frac23$, $\mu<0$), $p=2$ for the low hole concentrations
($2-n<\frac23$, $\mu>U$) with $p=\sigma,\bar\sigma$ for the
intermediate values. For the small electron or hole concentrations, the
Green's function for the atomic problem (\ref{eq36}) possess the
correct Hartree-Fock limits too. It was supposed that the Hubbard model
describes strongly-correlated electronic systems that contain four
components (subspaces). Subspaces $p=0$ and $p=2$ describe the
Fermi-liquid component (electron and hole, respectively) which is
dominant for the small electron and hole concentrations, when the
chemical potential is close to the bottom of the lower band and top of
the upper one (``overdoped regime'' of high-$T_c$'s). On the other
hand, subspaces $p=\uparrow$ and $\downarrow$ describe the
non-Fermi-liquid (strongly correlated, e.g., RVB) component, which is
dominant close to half-filling (``underdoped regime'').

\begin{figure}[htbp]
 \centerline{\includegraphics[width=.47\textwidth]{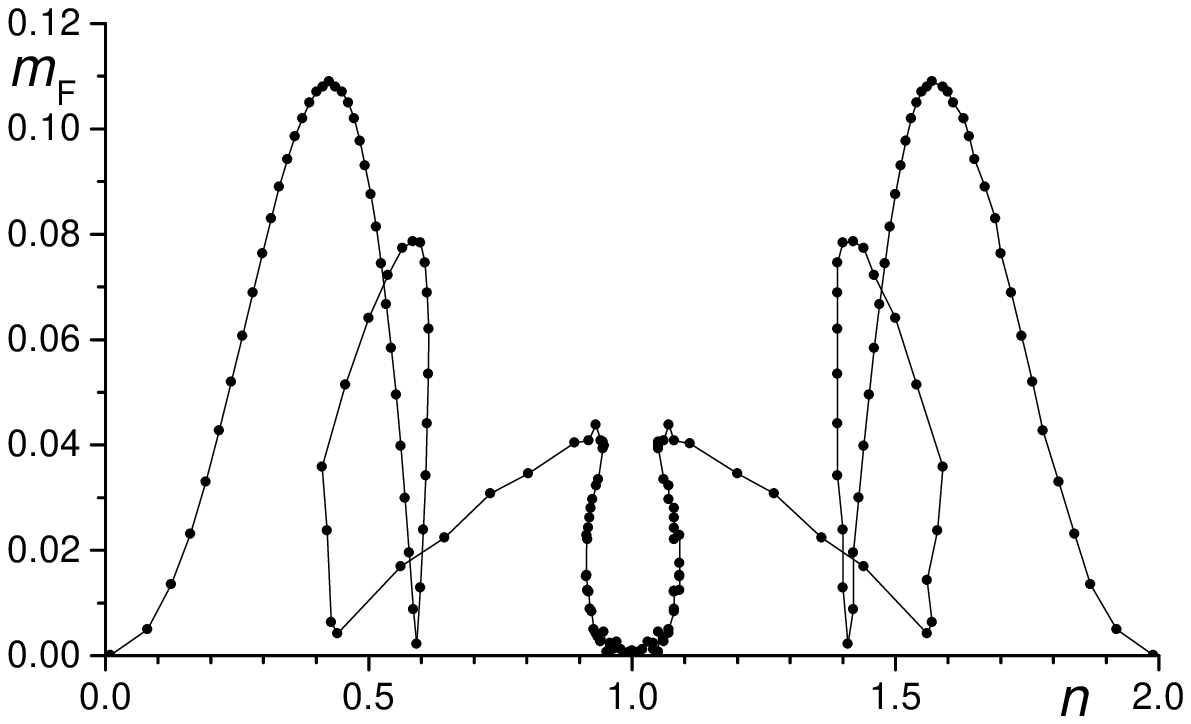}\quad
 \includegraphics[width=.47\textwidth]{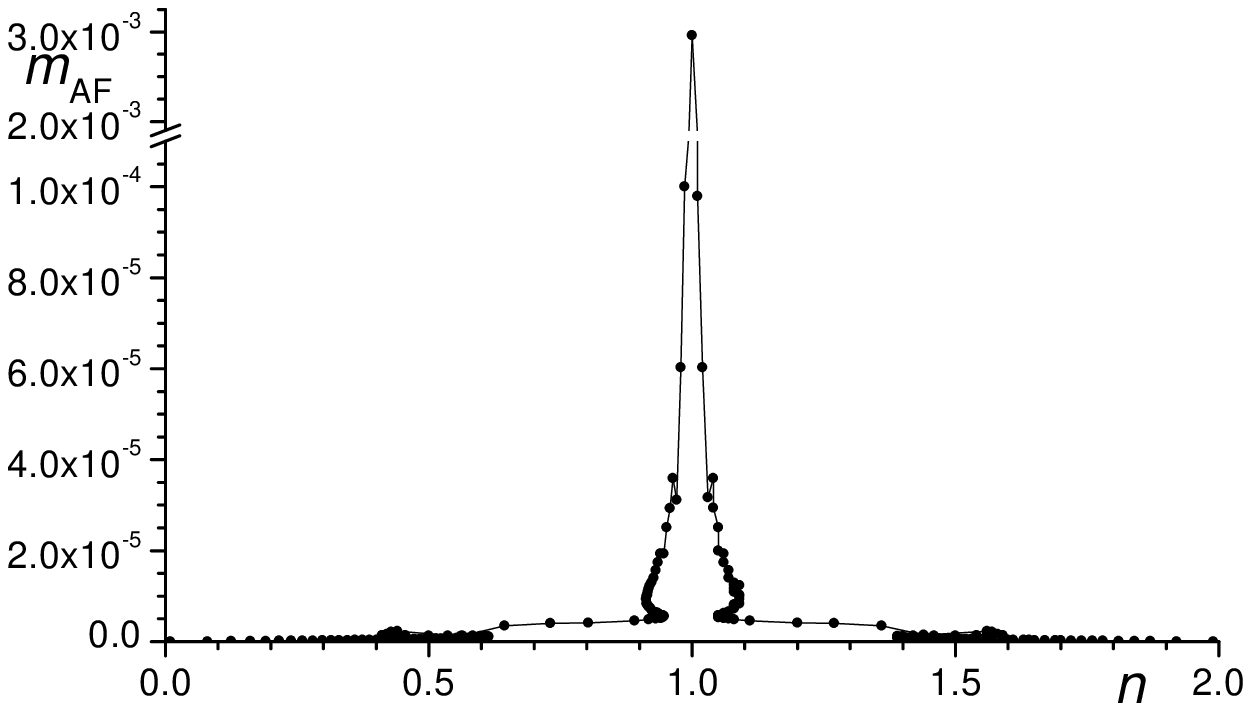}}
 \caption{Ferromagnetic $m_{F}$ and antiferromagnetic $m_{AF}$ order
 parameters vs electron concentration for $U=1.56$, $T=0.14$.}
 \label{FerrOP}
 \end{figure}

Here we consider possible magnetic orderings. At low temperatures the
$p=0$ and $p=2$ components for the low electron or hole concentrations
are in the ferromagnetic state, while the non-Fermi-liquid one is
antiferromagnetic (AF) close to half-filling, see Fig.~\ref{FerrOP}.
For the intermediate concentration values the picture is very
complicated, even frustrated. It is due to the fact that equations for
the mean values (\ref{mvHF}) have several solutions in this region,
which, on the other hand, are mutually connected with the dynamical
mean field $J_{\sigma}(\omega_n)$. It is difficult to determine the
ground state for this, possibly ``pseudo-gap'', region, which is
located between the ferromagnetic and antiferromagnetic phases.

\begin{figure}[htbp]
\centerline{\includegraphics[width=.7\textwidth]{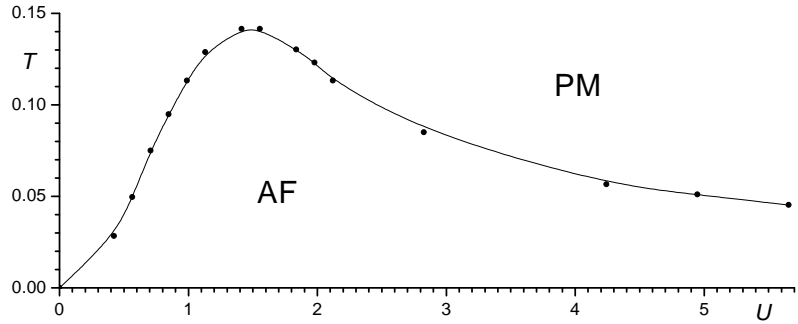}}
 \caption{Phase diagram $(T,U)$ at half-filling $n=1$ (AF~--- antiferromagnetic phase,
 PM~--- paramagnetic phase).}
 \label{T-U}
\end{figure}

In Fig.~\ref{T-U} we presented the phase diagram $(T,U)$~--- the
temperature of the AF ordering vs correlation energy $U$, which is in a
qualitative agreement with the results of Refs.~\cite{Pruschke,T-U} and
reproduces the results of the Hartree-Fock theory and mean field
approximation for $U\ll t$ and $U\gg t$, respectively. Our results for
the AF critical temperature for small $U$ are higher then the one of
the Quantum Monte Carlo simulations \cite{Pruschke} by about a factor
at three that describes the reduction of the Hartree-Fock solution by
the lowest order quantum fluctuations \cite{Dongen}.

\end{document}